\documentclass{emulateapj}

\usepackage{amsmath}
\usepackage{amssymb}
\usepackage{graphicx}
\usepackage{color}
\usepackage{natbib}

\def\gtaprx {\lower .1ex\hbox{\rlap{\raise .6ex\hbox{\hskip .3ex
	{\ifmmode{\scriptscriptstyle >}\else
		{$\scriptscriptstyle >$}\fi}}}
	\kern -.4ex{\ifmmode{\scriptscriptstyle \sim}\else
		{$\scriptscriptstyle\sim$}\fi}}}
\def\ltaprx {\lower .1ex\hbox{\rlap{\raise .6ex\hbox{\hskip .3ex
	{\ifmmode{\scriptscriptstyle <}\else
		{$\scriptscriptstyle <$}\fi}}}
	\kern -.4ex{\ifmmode{\scriptscriptstyle \sim}\else
		{$\scriptscriptstyle\sim$}\fi}}}

\newcommand{\cutt}[1]{\textcolor{blue}{}}

\begin{document}

\title{The Formation of Supermassive Black Holes from Low-Mass Pop III Seeds}

\author{Daniel J. Whalen\altaffilmark{1} and Chris L. Fryer\altaffilmark{2}}

\altaffiltext{1}{McWilliams Fellow, Department of Physics, Carnegie Mellon 
University, Pittsburgh, PA 15213}

\altaffiltext{2}{CCS-2, Los Alamos National Laboratory, Los Alamos, NM 87545}

\begin{abstract}

The existence of 10$^9$ M$_{\odot}$ black holes (BH) in massive galaxies by $z \
sim$ 7 is one of the great unsolved mysteries in cosmological structure formation.  
One theory argues that they originate from the black holes of Pop III stars at $z \sim$ 
20 and then accrete at the Eddington limit down to the epoch of reionization, which 
requires that they have constant access to rich supplies of fuel.  Because early 
numerical simulations suggested that Pop III stars were $\gtrsim$ 100 M$_{\odot}$, 
the supermassive black hole seeds considered up to now were 100 - 300 M$_{\odot}$. 
However, there is a growing numerical and observational consensus that some Pop III 
stars were tens of solar masses, not hundreds, and that 20 - 40 M$_{\odot}$ black 
holes may have been much more plentiful at high redshift.  However, we find that 
natal kicks imparted to 20 - 40 M$_{\odot}$ Pop III BHs during formation eject them 
from their halos and hence their fuel supply, precluding them from Eddington-limit 
growth.  Consequently, supermassive black holes are far less likely to form from 
low-mass Pop III stars than from very massive ones.

\end{abstract}


\maketitle

\section{Introduction}

The existence of 10$^9$ M$_{\odot}$ black holes (BH) in massive galaxies by 
$z \sim 7$, only a billion years after the Big Bang \citep[e.g.][]{mort11}, poses 
one of the great unsolved problems in cosmological structure formation.   In the 
$\Lambda$CDM paradigm, early structure formation is hierarchical, with small 
objects at high redshifts evolving into ever more massive ones by accretion and 
mergers through cosmic time. For this reason it is generally supposed that the 
supermassive black holes (SMBH) that power the $z \sim 7$ \textit{Sloan Digital 
Sky Survey} (\textit{SDSS})  quasars grow from much smaller seeds at earlier 
epochs.  The origin of SMBH and how they reach such large masses in such 
short times is a subject of ongoing debate.  Three modes of formation have been 
proposed for SMBH seeds: the collapse of Pop III stars into 100 - 300 M$_{\odot}$ 
black holes at $z \sim 20$ \citep{awa09}, baryon collapse in 10$^8$ M$_{\odot}$ 
dark matter halos that have somehow bypassed previous star formation into 
10$^4$ - 10$^6$ M$_{\odot}$ BH at $z \sim 15$ \citep{wta08,rh09,sbh10}, 
and more exotic pathways like the relativistic collapse of dense primeval star 
clusters into 10$^4$ - 10$^6$ M$_{\odot}$ BH \citep[see section 3.3 of][for a 
recent review]{brmvol08}.  

Stellar-mass SMBH seeds form at $z \sim 20$ when Pop III stars die in either 
core-collapse supernovae (SNe, 15 - 45 M$_{\odot}$) or by direct collapse to 
a BH (45 - 100 M$_{\odot}$, $\gtrsim 260$ M$_{\odot}$) \citep{hw02}.  This 
formation channel is favored by some because most dark matter halos will 
form a Pop III star at this epoch if they reach masses of $\sim$ 10$^5$ M$_
{\odot}$ \citep{abn02, bcl02}.  However, these BH have such low initial 
masses that they must continuously accrete at the Eddington limit to reach 
10$^9$ M$_{\odot}$ by $z \sim 7$.  This is problematic for several reasons. 
First, numerical simulations have shown that Pop III stars usually evaporate 
the halos that give birth to them, so the BH are 'born starving' \citep[e.g.][]{
wan04,ket04,wet08a}.  Filamentary inflows and mergers later restore baryons 
to the halo but only after 50 - 100 Myr \citep{yet07}, during which crucial 
e-foldings in mass are lost.  Second, preliminary studies indicate that once 
accretion commences, the BH itself emits ionizing radiation that disperses its 
own fuel supply, limiting its growth rate to a fraction of the Eddington limit \citep{
milos09,pm11,pm12} \citep[but see][]{li11}. Furthermore, if the seed BH is not 
confined to the halo, its duty cycle as it meanders through cosmological density 
fields is intermittent, which also curtails its growth \citep{awa09}.  

Until now, 20 - 40 M$_{\odot}$ Pop III BH \citep{zwh08,wet08b} have been 
overlooked as candidates for SMBH seeds because previous studies assume 
that primordial stars are $\gtrsim$ 100 M$_{\odot}$.  However, there is a growing 
numerical and observational consensus that some Pop III stars are tens of solar 
masses, not hundreds.  More recent, much larger ensembles of numerical 
simulations found many halos with central collapse rates consistent with 20 - 60 
M$_{\odot}$ for the final mass of the star \citep{on07} and that a fraction of the 
halos form binaries in this mass range \citep{turk09}.  Furthermore, new 
simulations of the formation of Pop III protostellar accretion disks suggest that 
they were prone to fragmentation into as many as a dozen smaller stars \citep{
stacy10,clark11,sm11,get11}.  Very preliminary calculations of I-front breakout 
from these disks indicate that ionizing UV radiation may terminate accretion onto 
the nascent star at $\sim$ 40 M$_{\odot}$ \citep{hos11,stacy12}. 

On the observational side, recent attempts to reconcile the nucleosynthetic yields 
of Pop III supernovae with the chemical abundances found in ancient, dim 
extremely metal-poor stars in the Galactic halo suggest that 15 - 40 M$_{\odot}$ 
primordial stars may have been responsible for most of the heavy elements 
expelled into the primeval IGM \citep{jet09b}.  The failure to detect the distinctive 
'odd-even' nucleosynthetic signature of 140 - 260 M$_{\odot}$ pair-instability SNe 
in metal-poor stars to date reinforces the fact that some Pop III stars might not be 
very massive, but this pattern may have been masked by selection effects in the 
observations \citep{karl08}.

Low-mass Pop III BH are crucially different from more massive BH because they 
are born in supernova explosions rather than by direct collapse.  Asymmetries in 
the core-collapse engine can impart kicks of 200 - 1000 km/s to 20 - 40 M$_{
\odot}$ BH, ejecting them from the halos that gave birth to them.  In this Letter 
we examine the implications of natal kicks for low-mass Pop III black holes as 
candidates for SMBH seeds.  In $\S \, 2$ we review the formation pathways for 
low-mass Pop III BH.  In $\S \, 3$ we calculate their post-supernova kinematics 
and retention fractions in halos.  In $\S \, 4$ we conclude.

\section{Low-Mass Pop III Black Holes}

Three mechanisms can create Pop III black holes during stellar collapse. In order 
of increasing progenitor mass, they are fallback onto a neutron star (NS) during a 
supernova explosion, the direct collapse of a proto-neutron star into a BH without 
an explosion, and enclosure of the core by an event horizon without ever having 
attained nuclear densities \citep{fwh01,oco11}.  

\subsection{Pop III BH Formation}

It is generally believed that core-collapse supernova explosion energies fall with 
increasing progenitor mass \citep[see][for a review]{fryer03}.  At some point, the 
explosion is too weak to fully overcome the binding energy of the star and enough 
ejecta falls back onto the NS to collapse it to a black hole \citep{fcp99,zwh08}.  In 
even more massive progenitors, the core of the star collapses to a proto-neutron 
star without an explosion.  About 1\,s after the onset of collapse, it gains so much 
additional mass that it cannot support itself and it collapses to a black hole. In the
most massive stars ($\gtrsim$ 300\, M$_\odot$) the entropy of the core becomes 
so high that it never reaches nuclear densities.  When enough material falls into 
the core it is suddenly engulfed by an event horizon, forming a BH of $> 20$ M$_{
\odot}$ \citep{fwh01}.  In general, the birth masses of Pop III BH vary from the 
minimum black hole mass ($\sim 2-3 \, M_\odot$) up to the mass of the progenitor 
star.  

\subsection{Pop III BH Kicks}

The first two formation processes can impart kicks (initial velocity pulses) to low-mass 
BH at birth.  Kick mechanisms generally fall into two categories:   ejecta-driven kicks 
\citep[e.g.][]{janke06, janke10} and neutrino-driven kicks~\citep[see][and references 
therein]{kus06}.  Ejecta kicks occur when low-mode instabilities erupt in the shock as
it is driven outward by core bounce.  They are likely seeded during collapse prior to 
bounce and result in explosion asymmetries that impart a net linear momentum to the 
neutron star (NS).  Above 32 M$_{\odot}$ fallback is total, and no impulse is imparted 
to the BH.  Neutrino kicks arise when magnetic field lines through the center of the star 
are crushed to extremely high densities during collapse, polarizing neutrinos created 
by the core during deleptonization and inducing anisotropies in emission that deliver 
an impulse to the NS.  Consequently, neutrinos can impart momentum to the NS (and 
therefore the BH) even if there is no explosion.  

\begin{figure}
\epsscale{1}
\plotone{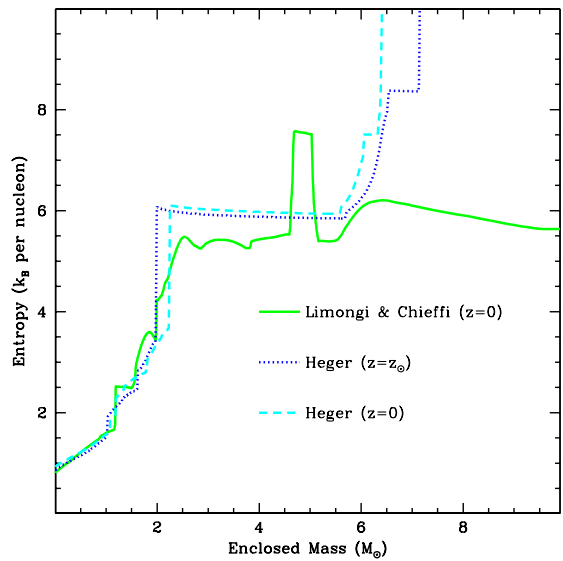}
\caption{Entropy versus enclosed mass for three 25\,M$_\odot$ stars.  Note that 
the differences due to the codes are greater than those due to metallicity. 
\vspace{0.1in} 
\label{fig:entropy} } 
\end{figure}

Pop III core-collapse SNe imparted kicks to neutron stars and black holes in the same
manner as in the Galaxy today.  Both kick mechanisms arise from asymmetries in the 
explosion engine that are determined by the structure of the inner core (inner 3-4\,M$
_\odot$) of the star.  We show the entropy profile of this core for three 25\,M$_\odot$ 
stars at collapse in Figure~\ref{fig:entropy}: a zero metallicity star modeled by \citet{cl04} 
which collapsed with a mass of 24.7\,M$_\odot$, a zero metallicity star modeled by 
\citet{wh07} which collapsed with a mass of 24.9\,M$_\odot$ and a solar metallicity star 
modeled by \citet{wh07} which collapsed with a mass of 12.9\, M$_\odot$.  Since 
metallicity has very little effect on the structure below 6\,M$_\odot$, the engine will not 
differ significantly between a zero and solar metallicity star and they will exhibit similar
kick distributions.  The structures of the cores of very massive stars do change with 
metallicity but we do not expect kicks in their supernovae.

\begin{figure*}
\epsscale{1.17}
\plottwo{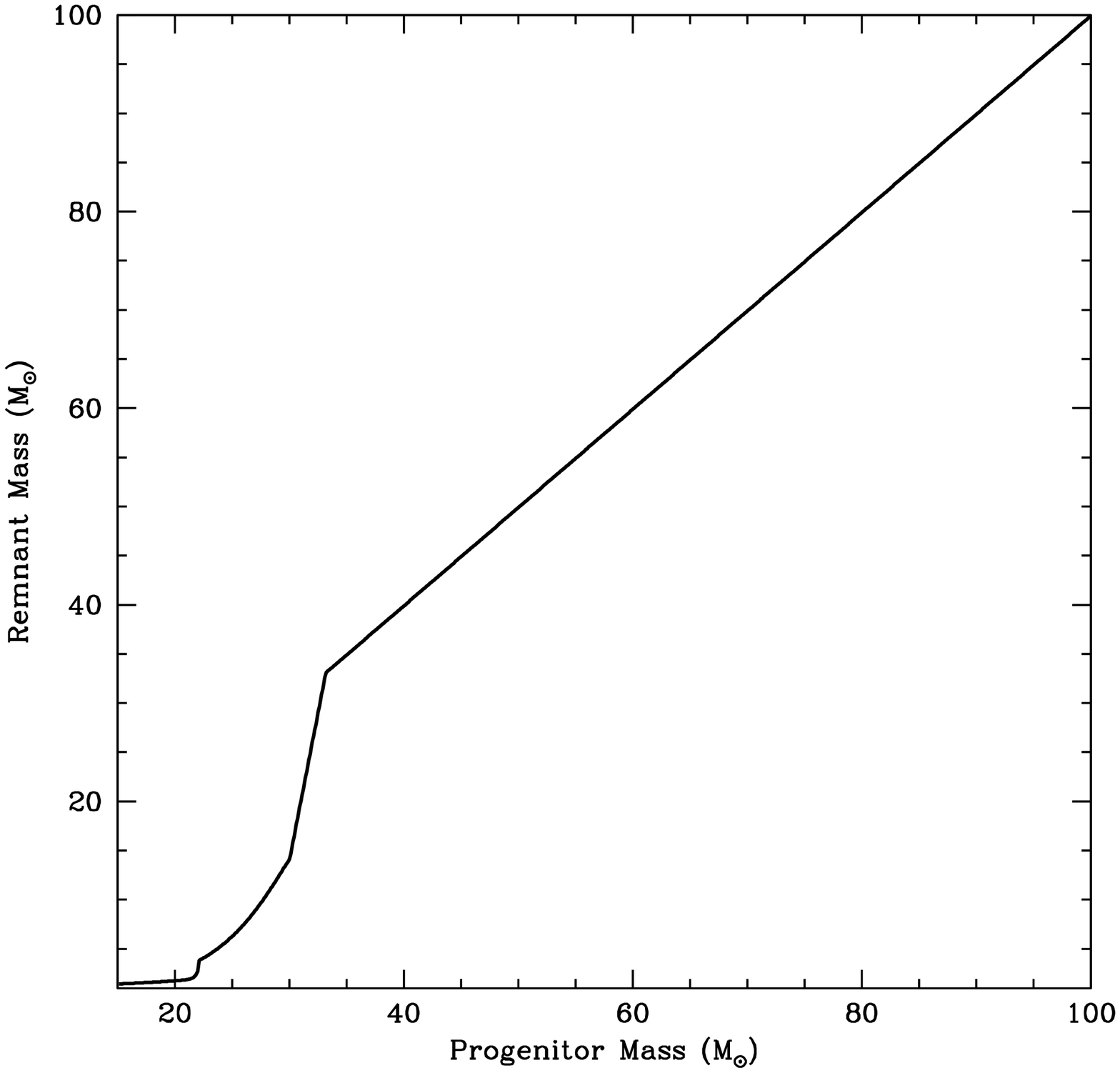}{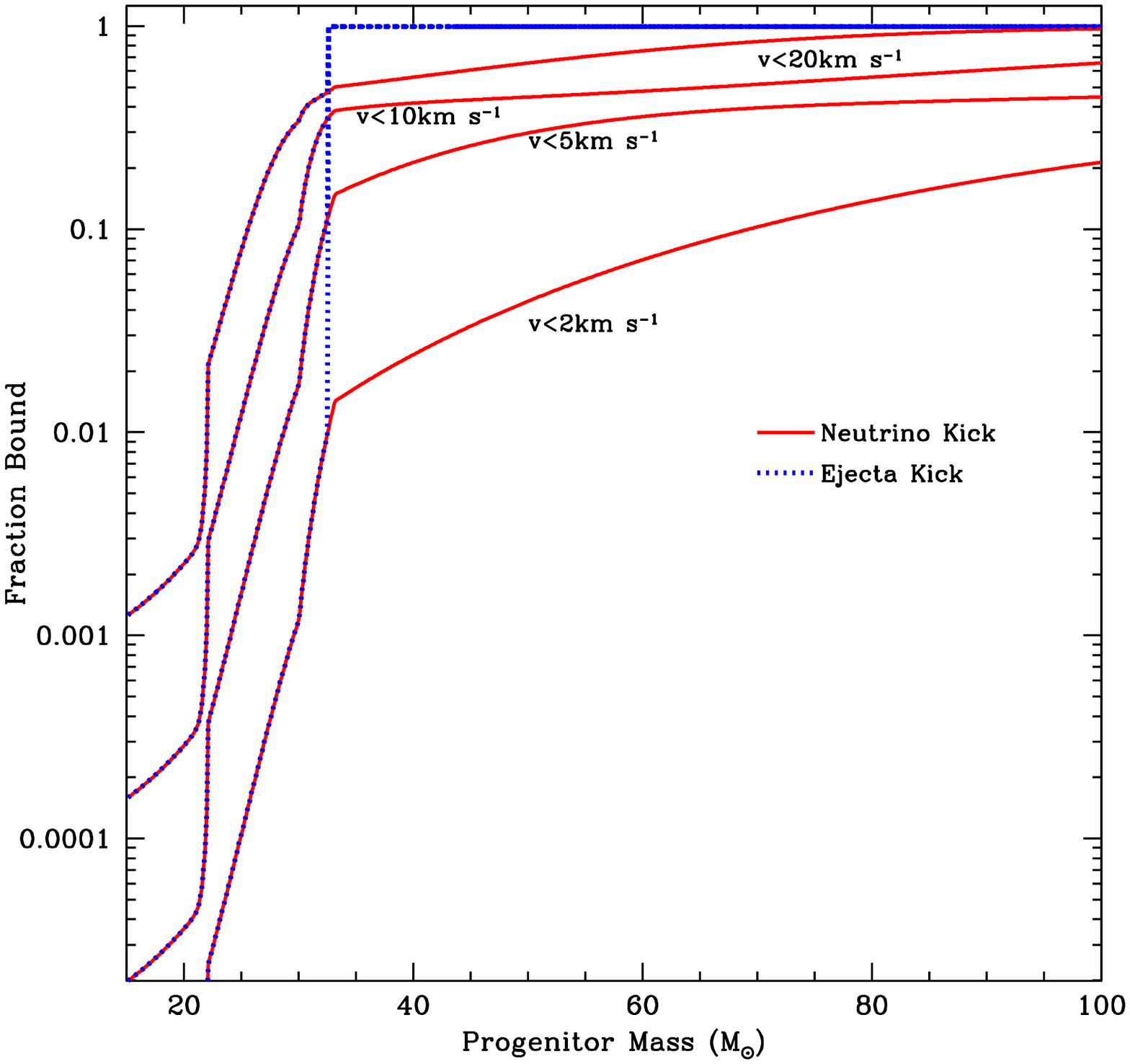}
\caption{Statistical properties of low-mass Pop III BH at birth.  Left:  black hole 
mass as a function of progenitor mass.  Right:  BH retention fraction in the halo 
as a function of progenitor mass.  \vspace{0.1in} \label{fig:BH_MR}}  
\end{figure*}

\section{Post-Supernova Kinematics of Low-Mass Pop III BH}

Fully developed models for kick mechanisms do not yet exist, so neither the number 
of Pop III seed BH kicks nor their velocity distributions can be calculated from first 
principles.  However, natal kicks are commonly observed in compact remnants in the 
Galaxy today and there are models that infer reasonable relationships between BH 
and NS kick distributions, which have been measured for a large sample of pulsars. 
In our study we adopt the pulsar velocity distribution of  \citet{arz02}.  It is bimodal, 
with each mode being described by a Maxwellian:  40\% have a dispersion of 90 km/s
and 60\% have a dispersion of 500\,km\,s$^{-1}$.  We derive velocity distributions 
for low-mass Pop III BH by assuming that in both mechanisms the black hole simply 
inherits the linear momentum of the NS: \vspace{0.075in}
\begin{equation}
v_{BH} = v_{NS} \frac{m_{NS}}{m_{BH}},  \vspace{0.075in}
\end{equation}
where $m_{NS}$ is the Chandrasekar mass, 1.4 M$_{\odot}$.  Consequently, the BH 
kick velocity is inversely proportional to its mass.  We derive our BH mass distribution 
from the latest estimates of \citet{fh11} for zero-metallicity stars, assuming rapid 
explosions.  We show the distribution for these new fits in the left panel of Figure 
\ref{fig:BH_MR}.  In reality, the BH could acquire more momentum than the NS 
intermediary because of the tendency of weak explosions to be more delayed, which 
allows low-mode instabilities additional time to develop and create greater asymmetry 
in the ejecta \citep{fh11}.  With our black hole mass and pulsar velocity distributions we 
can estimate the retention fraction of BH in halos as a function of progenitor mass for a 
variety of escape velocities from the halo, as we show in the right panel of 
Figure~\ref{fig:BH_MR}.  Above $\sim 32\,M_\odot$, the kick velocity drops to zero for 
the ejecta mechanism and we expect full retention for stars above this mass limit in the 
absence of neutrino kicks.  If there are neutrino kicks, retention fractions for BH below 
40 M$_{\odot}$ are less than 10\% in the halos in which most Pop III stars form (those 
with $v_{esc} <$ 5 km/s), and fall below 1\% for BH below 32 M$_{\odot}$.

\section{Discussion and Conclusion}

The number of stars that 10$^5$ - 10$^7$ M$_{\odot}$ halos typically form is not well 
constrained.  The studies of Pop III protostellar disk fragmentation performed thus far 
do not follow the evolution of the disk for enough dynamical times to determine the 
ultimate fate of the fragments, which may later merge with the central object or be 
destroyed by gravitational torques before becoming distinct stars.  Ionizing UV 
radiation from one star-forming fragment or even from a nearby halo can also
prematurely halt the collapse of other fragments in the disk, lowering the number of 
stars that eventually form in the halo \citep[e.g.][]{su06,wet08b,suh09,wet10}.  We also 
note that while the evolution of the fragments in the disk is expected to be roughly 
coeval, their 5 - 10 Myr quasistatic collapse times raise the possibility that the first star 
to form in the halo may explode and pre-empt the collapse of other fragments 
\citep[e.g.][]{sak09}.  Consequently, the number of low-mass Pop III stars that occupy 
the halo likely ranges from one to at most ten.  

Ejecta-driven natal kicks will evict most 20 - 32 M$_{\odot}$ BH from their host halos, 
neutrino-driven kicks can drive more than 90\% of 32 - 40 M$_{\odot}$ BH from their 
halos, as we show in the right panel of Figure \ref{fig:BH_MR}.  This guarantees that 
on average all the BH will vacate the halo even if ten stars originally formed in it.  
Post-supernova kinematics thus strongly discourages 20 - 40 M$_{\odot}$ Pop III BH 
from becoming supermassive because they are ejected from their fuel supply and
deprived of crucial early e-foldings in mass.  This process greatly reduces the 
parameter space in stellar mass from which SMBH can originate \citep[e.g.][]{th09,
lfh09}, especially if Pop III stars were mostly less than 50 M$_{\odot}$. Also, if a given 
halo is capable of supporting early continuous Eddington rate accretion, a 20 - 40 M$
_{\odot}$ BH is much less likely to become supermassive than a 100 M$_{\odot}$ BH, 
either because it is ejected from the halo at birth or because it must undergo additional 
e-folding times to reach large masses.  

If most low-mass Pop III black holes were ejected from their halos at $z \sim$ 20, where
are they today?  If on average they depart their host halos at $\sim$ 500 km/s, they are
unlikely to encounter another halo capable of capturing them in less than a Hubble time,
and so many of these BH were exiled to the voids between galaxies. Over time, they may 
have gradually gained mass as they encountered high-density regions.  In contrast, Pop 
III BH above 40 M$_{\odot}$ are unlikely to be born with kicks and remain in the halo, 
intermittently accreting and growing over cosmic time. These black holes are much more 
likely to reside in the galaxies into which their host halos were taken, a few of which may 
have become the supermassive black holes found in the SDSS quasars today.

\acknowledgments

We thank the anonymous referee for comments that improved the quality of this paper  
and Jarrett Johnson and Brian O'Shea for valuable comments.  DJW was supported by 
the Bruce and Astrid McWilliams Center for Cosmology at Carnegie Mellon University.  
Work at LANL was done under the auspices of the National Nuclear Security 
Administration of the U.S. Department of Energy at Los Alamos National Laboratory under 
Contract No. DE-AC52-06NA25396.

\end{document}